\documentclass[amsmath,amssymb,aps,mathbbm,superscriptaddress,twocolumn,notitlepage]{revtex4-1}
\usepackage{amsmath,amsfonts,amssymb,amsthm,graphics,graphicx,epsfig,times,bbm}

\usepackage[colorlinks=true,citecolor=blue,linkcolor=magenta]{hyperref}
\usepackage[usenames]{color}

\definecolor{nils}{rgb}{0,0,0}
\definecolor{jens}{rgb}{0,0,0}
 
\newcommand{\je}{\textcolor{black}} 
\newcommand{\jens}{\textcolor{black}}

\newcommand{\rr}{\mathbbm{R}}

\begin{document}

\title{
Observation of non-Markovian micro-mechanical Brownian motion
}

\author{S.\ Gr\"oblacher}
\affiliation{Kavli Institute of Nanoscience, Delft University of Technology, 2628 CJ Delft, Netherlands}
\affiliation{University of Vienna, Vienna Center for Quantum Science and Technology (VCQ), Faculty of Physics, A-1090 Vienna, Austria}
\author{A.\ Trubarov}
\affiliation{University of Vienna, Vienna Center for Quantum Science and Technology (VCQ), Faculty of Physics, A-1090 Vienna, Austria}
\author{N.\ Prigge}
\affiliation{Dahlem Center for Complex Quantum Systems, Freie Universit{\"a}t Berlin, 14195 Berlin, Germany}
\author{G.\ D.\ Cole}
\affiliation{University of Vienna, Vienna Center for Quantum Science and Technology (VCQ), Faculty of Physics, A-1090 Vienna, Austria}
\author{M.\ Aspelmeyer}
\affiliation{University of Vienna, Vienna Center for Quantum Science and Technology (VCQ), Faculty of Physics, A-1090 Vienna, Austria}
\affiliation{Dahlem Center for Complex Quantum Systems, Freie Universit{\"a}t Berlin, 14195 Berlin, Germany}
\author{J.\ Eisert}
\affiliation{Dahlem Center for Complex Quantum Systems, Freie Universit{\"a}t Berlin, 14195 Berlin, Germany}

\date{\today}

 \begin{abstract}
\jens{All physical systems are to some extent open and interacting with their environment. This insight, basic as it may seem, gives rise to the necessity of protecting quantum systems from decoherence in quantum technologies and is at the heart of the emergence of classical properties in quantum physics. The precise decoherence mechanisms, however, are often unknown for a given system. In this work, we make use of an opto-mechanical resonator to obtain key information about spectral densities of its condensed-matter heat bath. In sharp contrast to what is commonly assumed in high-temperature quantum Brownian motion describing the dynamics of the mechanical degree of freedom,
based on a statistical analysis of the emitted light, it is shown that this spectral density is highly non-Ohmic, reflected by 
non-Markovian dynamics, which we quantify. We conclude by elaborating on further applications of opto-mechanical
systems in open system identification.}
 \end{abstract}
\maketitle
\medskip

\section{Introduction}

At the heart of understanding the emergence of a classical world from quantum theory is the insight that all macroscopic quantum systems are to some extent coupled to an environment and hence are open systems~\cite{Joos1996,Zurek2003,Gardiner2004,Weiss}. The associated loss of quantum coherence, i.e.\ decoherence, is also detrimental for quantum information processing applications. In contrast, properly engineered quantum noise can counteract decoherence and can even be used in robust quantum state generation~\cite{Cirac,Diehl2008,Eisert2010}. To exploit the detailed dynamics of a quantum system it is therefore crucial to obtain both good knowledge and control over its environment~\cite{Loss,Marcus,DiFranco}. An explicit modelling of the environment, however, may often not be possible. In this case, simplifying assumptions concerning the nature of the underlying quantum noise are being made that do not necessarily hold for real devices. Micro- and nano-mechanical resonators constitute prominent examples. They are now emerging as promising devices for quantum science~
\cite{Cleland2004,Rabl2010,Martinis,Chan2011,Riviere2010,Weis2010,Teufel}.
Because of their complex solid-state nature, the properties of their intrinsic decoherence mechanisms 
have been the subject of intense research for decades~\cite{Wilson-Rae2008b,Kotthaus2010}. 

In this work, we present a method to reconstruct the relevant properties of the environment, that is, its spectral density, of the centre of mass motion of a micro-mechanical oscillator. We observe a clear signature of non-Markovian Brownian motion,  which is in contrast to the current paradigm to treat the thermal environment of mechanical quantum resonators as fully Markovian. The presented technique, inspired by methods of system identification, can easily be transferred to other harmonic systems that are embedded in a complex environment, for example electronic or nuclear spin states in a solid state matrix~\cite{qdots,NV}. Our results also open up a route for mechanical quantum state engineering via coupling to unorthodox reservoirs.

\section{Results}

\subsection{\jens{Open quantum systems}}
To understand the role of the environment on a (quantum) mechanical system let us first consider an isolated harmonic oscillator of bare frequency $\Omega$ and mass $m$. In the absence of any coupling its centre of mass coordinate $q$ will undergo undamped harmonic motion. In any real physical situation, however, the macroscopic degree of freedom of interest -- here the centre of mass -- will be coupled to some extent to a thermal bath of some temperature. Irrespective of the underlying microscopic mechanism, e.g., phonon scattering in mechanical systems \cite{Lifshitz} or electronic interactions in superconductors~\cite{Leggett1984}, one can usually very well approximate the interaction with the thermal environment as a linear coupling to a bath of harmonic bosonic modes~\cite{Feynman1963}. This is particularly true for high temperatures where finite bath degrees of freedom no longer significantly contribute. Such an interaction is described by 
\begin{equation}
H_{\rm int}=q\sum_n c_n q_n 
\end{equation}
where $q_n$ and $c_n$ are the position and coupling strength of the $n^{\rm th}$ bath mode of mass $m_n$ and frequency $\omega_n$, respectively. The dynamics of the system is fully determined by the spectral density of the thermal bath,
\begin{equation}
I(\omega)=\sum_n \frac{c_n^2}{2m_n\omega_n}\delta(\omega-\omega_n),
\end{equation} 
which governs how strongly the oscillator is coupled to specific modes of the environment. This spectral density directly determines the temporal correlations of the thermal driving force. As a consequence, the centre of mass experiences a quite drastic change in its motion: it becomes damped, in general in a rather intricate fashion, and is shifted in its frequency. This {\it quantum Brownian motion}~\cite{Caldeira1983,Hu1992} is one of the most paradigmatic models of decoherence in quantum theory~\cite{Joos1996,Zurek2003,FlemingHu,EisertPlenio}. 
\je{It is this generic model for an unknown arbitrary spectral density that is the basis for our analysis.}

All current theoretical studies on micro- and nano-mechanical quantum systems make the explicit or implicit assumption that the decohering quantum dynamics is Markovian: This means that the open systems dynamics is forgetful \cite{Markovian1,Markovian3,Markovian2,Maniscalco}. In this case the two-point correlation function of the thermal force
equals $k_B T\delta(t-t')$ and is hence uncorrelated in time. 
For a weakly damped mode at high temperatures
(and in contrast to the situation in spin-Bose models \cite{DiVincenzo}),
such Markovian quantum dynamics is found for an {\it Ohmic spectral density}
\begin{equation}
I(\omega) \propto \omega 
\end{equation}
over large frequency ranges. For such damped harmonic systems in the high temperature limit, 
spectral densities other than Ohmic ones lead to deviations from Markovian evolutions. \je{This is a widely
known expectation  \cite{Hu1992,OConnell2010,FlemingHu}. In this work we precisely link properties of spectral 
densities with a quantitative measure of non-Markovianity.}

In many solid-state systems the Markov approximation has been found to be both theoretically plausible and experimentally valid to extraordinarily high precision~\cite{Gardiner2004}. Various loss mechanisms in mechanical resonators, however, are known to exhibit a strong frequency dependence~\cite{Wilson-Rae2008b,Lifshitz}, which challenges the general validity of this approximation even for simple mechanical quantum devices. We introduce a straight-forward test to directly characterise the spectral properties of the environment in the vicinity of the mechanical mode. Because of the complex solid-state architecture of these resonators, computing the spectral density from first principles seems a tedious if not impossible task, with the exception of well isolated loss mechanisms. Instead of making a priori assumptions about the dynamics, our approach is rather in the spirit of {\it open system identification}:\ we measure the properties that give rise to a quantitative estimate on the Markovian nature of the dynamics. 

\begin{figure}[t]
\centering
\includegraphics[width=1\linewidth]{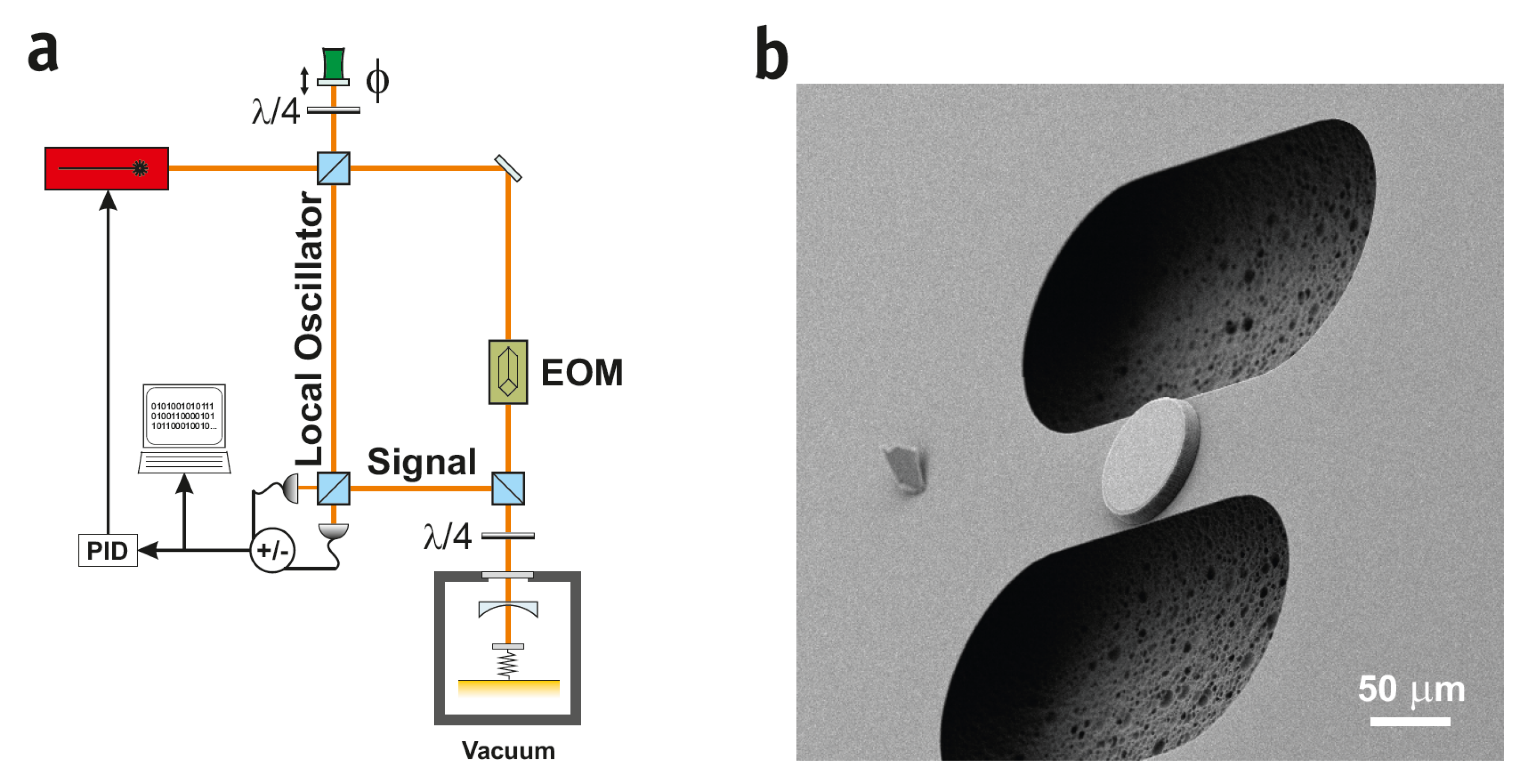}
\caption{\textbf{Sketch of the experiment.} \textbf{a} The experimental setup consist of a 1064~nm Nd:YAG laser, which is split into a signal beam and a local oscillator (LO). The signal is phase modulated with an electro-optical modulator (EOM) for Pound-Drever-Hall locking of the opto-mechanical cavity. In order to read-out the phase of the signal beam acquired from the motion of the mechanical resonator, it is beaten with a strong local oscillator (LO) on a beamsplitter and detected on two photodiodes. The phase $\phi$ between the LO and the signal is stabilised with the help of a mirror mounted on a piezo-ceramic actuator in order to only detect the phase quadrature of the signal field. The opto-mechanical cavity is kept at a pressure of $<10^{-3}$~mbar to avoid residual-gas damping of the mechanical motion. \textbf{b} Scanning electron microscope picture of the tested device.}
\label{Fig:1}
\end{figure}

\subsection{\jens{Experimental set-up}}

Our approach relies on monitoring the mechanical motion with high sensitivity. We achieve this by weakly coupling the mechanics to an optical cavity field whose phase response encodes the mechanical motion \cite{OptomechanicsRMP}. 
We then make use of the fact that the shape of the bath spectral density affects the amplitude response of the mechanical resonator upon thermal driving. Specifically, the experimentally accessible spectrum of the cavity output light for high temperatures is given by 
\begin{equation}\label{Spectrum}
S_{\delta Y^{\rm out}}(\omega) \approx c\frac{I(\omega)}{\omega\left((\Omega(\infty)^2-\omega^2)^2 + (\gamma(\infty)\omega)^2\right)},
\end{equation}
for a suitable constant $c>0$ (for details, see \jens{note 2 of}
the supplementary material). Here, $\delta Y^{\rm out}$ is the optical phase quadrature, which can be made a direct measure of the mechanical position quadrature $q$ and which is obtained by optical homodyne readout, 
$\Omega(\infty)$ is the renormalised mechanical frequency, and $\gamma(\infty)$ is the effective asymptotic mechanical damping constant. The opto-mechanical device can hence be seen as an ultra-sensitive {\it black box} measuring the spectral density.

\begin{figure}[t]
\centering
\includegraphics[width=0.95\linewidth]{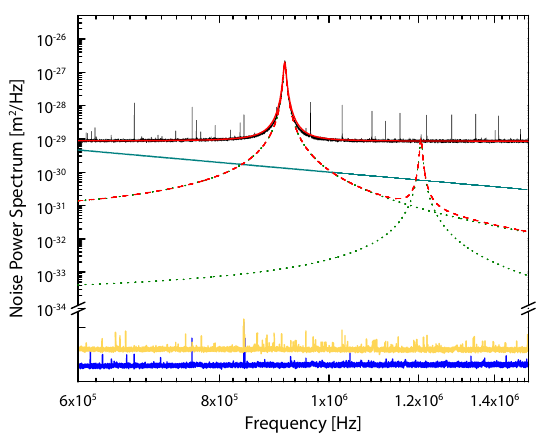}
\caption{\textbf{Noise power spectra.} Depicted are the spectra obtained with the mechanics being part of the setup (black) (with a fit in red), with no mechanics (yellow), with no cavity (blue), a spectrum reflecting a sub-Ohmic spectral density $I(\omega)\propto \omega^{-2}$ (turquoise), the simulated sum (red dashed) and the simulated modes (green dotted). {In our simulation we have assumed the mechanical Qs of the higher order modes to be similar to the fundamental mode, which is in good agreement with typical experimental values. Note that for clarity the measurements of the additional noise (yellow and blue) are not to scale.}}
\label{Fig:2}
\end{figure}

We demonstrate our analysis on a micro-mechanical resonator as shown in Figure~\ref{Fig:1}b. The device consists of a 1~$\mu$m thick layer of Si$_3$N$_4$ and is 150~$\mu$m long and 50~$\mu$m wide. The 50~$\mu$m diameter, high-reflectivity ($R>99.991\%$) mirror pad in its centre allows to use this resonator as a mechanically moving end mirror in a Fabry-P\'{e}rot cavity, as has been fabricated to explore the regime of cavity opto-mechanical coupling~\cite{Groeblacher2009a,Groeblacher2009b} (for details on the fabrication process see Ref.~\cite{GroeblacherPhD}). In our case, the cavity finesse is intentionally kept low at $F=2300$ by choosing a high-transmittivity input mirror
for this experiment. 
This results in an amplitude cavity decay rate of $\kappa=1.3$~MHz (cavity length: 25~mm). By using a signal beam of $100$~$\mu$W, we realise a sufficiently weak opto-mechanical coupling $g~\text{kHz}\ll\kappa$, such that the cavity field phase quadrature adiabatically follows the mechanical motion and hence $\delta Y^{\rm out}$ is a reliable measure of $q$. The fundamental mechanical resonance frequency is $\Omega=2\pi\times 914$~kHz, with a mechanical quality $Q$-factor of approximately $215$ at room temperature. Optical homodyne detection of the outgoing cavity field finally yields the temporal phase quadrature fluctuations $\delta Y^{\rm out}(t)$, which are digitised to calculate the  noise power spectrum $S_{\delta Y^{\rm out}}(\omega)$ (see Fig.~\ref{Fig:1}a). All experiments have been performed in vacuum (background pressure $<10^{-3}$~mbar) to prevent the influence of fluidic damping. {At the mentioned parameters for our experiment we achieve a displacement sensitivity of approx.\ 
$3\times10^{-15} {\mathrm m}/\sqrt{{\mathrm Hz}}$
as is shown in Fig.~\ref{Fig:2}}. To exclude the possible influence of spurious background noise we have also characterised the noise power spectrum of the cavity field without a mechanical resonator. In our configuration this is possible due to the specific design of the chip comprising the micro-mechanical device, which holds several non-suspended mirror pads which can be accessed by translating the chip. The resulting noise power spectrum is flat and hence cavity noise cannot contribute to any non-Brownian spectral signal (see Fig.~\ref{Fig:2}). Another possible spectral dependence could arise from the presence of higher-order mechanical modes, which are not taken into account in Eq.~(\ref{Spectrum}). A finite element analysis of our mechanical system reveals the next mechanical mode at $\Omega^{(1)}=2\pi\times 1.2$~MHz. As can be seen from Fig.~\ref{Fig:2}, the spectral overlap in the vicinity of $\Omega$ is many orders of magnitude below the measured signal and hence negligible.

\subsection{\jens{Spectral densities and non-Markovian dynamics}}

After characterising the resonator, the final task to perform bath spectroscopy now reduces to assessing the statistical significance of a single assumption: namely that the spectral density is locally, i.e.\ in the vicinity of 
\je{an estimate of} $\Omega$, well described by
\begin{equation}\label{OptimalK}
I(\omega) = C \omega^k,
\end{equation}
for some $C>0$ and $k\in \rr$, \je{for $\omega\in [\omega_{\rm min}, \omega_{\rm max}]$}. A value of $k=1$ corresponds 
to an {\it Ohmic environment}, $k>1$ to a {\it supra-Ohmic}, and $k<1$ to a {\it sub-Ohmic} environment. 
This is the common classification of spectral densities~\cite{Hu1992}. For a slowly varying spectral density, however,
what largely determines the long time dynamics is the slope of the spectral density in the vicinity of $\Omega$. 
\je{Indeed, for this analysis to be valid, we do not have to make a global model for the spectral density -- information that is experimentally inaccessible anyway -- 
but merely for the local frequency dependence. We accompany this analysis with an analytical assessment in
\je{notes 2 and 3 of} the supplementary material. 
}

The starting point of this analysis is Eq.~(\ref{Spectrum}). From the homodyne measurement, samples of statistically independent subsets of time series are formed, and data sets are obtained as Fourier transforms thereof. For each of these independently distributed Fourier transforms, one identifies the optimal $k$ in Eq.~(\ref{Spectrum}) with $I(\omega)= C \omega^k$ that minimises the least square deviation
within a suitable frequency interval ${[}\omega_{\rm min}, \omega_{\rm max}{]}$
centred around $\Omega$. Here $\omega_{\rm min}= 885$~kHz and $\omega_{\rm max}= 945$~kHz are chosen, but the results are largely independent of that choice. Interestingly, it is the comparably low mechanical $Q$-factor that allows for the assessment of a relatively large frequency interval. For each individual data set, several different values of the power $k$ are compatible with the data, which is 
an unsurprising finding in the light of the presence of noise in the data. Given the large data set that is available, however, one can arrive at an estimate of the optimal coefficient $k$ with large statistical significance. 

The main experimental result is shown in Fig.~\ref{Fig:histogram} \jens{(see also note 6 of the supplementary material)}. The histogram over all optimal power estimates yields $k=-2.30\pm 1.05$, which is a clear deviation from $k=1$ for a locally Ohmic bath density, hence signifying 
a remarkably strong departure from Markovianity.
\je{It is well known that an Ohmic spectral density leads in the weak coupling and high temperature regimes to Markovian dynamics \cite{OConnell2010,FlemingHu}. To further strengthen 
our analysis, we further make this link quantitative: 
We show that a deviation from a local Ohmic spectral density -- which is precisely what is observed -- leads to 
quantifiable non-Markovian dynamics.}

\subsection{\jens{Quantifying non-Markovian harmonic dynamics}}

\je{Formally, open systems dynamics is precisely Markovian if the time evolution is captured by
$\dot\rho(t) = {\cal L}(\rho(t))$,
with ${\cal L}$ being a Liouvillian.  
In order for it to give rise to a valid quantum channel and hence to
quantum dynamics, it has to take the so-called Lindblad form, 
\begin{equation}
	{\cal L}(\rho) =\sum_j\left(
	L_j \rho L_j^\dagger - \frac{1}{2}\{L_j^\dagger L_j,\rho\}
	\right).
\end{equation}
Obviously, any conceivable dynamics that is not
generated by Hamiltonian evolution will only be approximately Markovian. 
This approximation can, however, be exceedingly good. The channels resulting from
Markovian dynamics are infinitely divisible \cite{Markovian1,Divisibility}.}
\je{For harmonic systems, the exact master equation governing time evolution is of the form
\begin{align}
\label{HPZequation}
	\dot\rho(t) =&-i\left[H_{R}(t),\rho\right]-i\gamma (t)\left[x,\left\{ p,\rho(t) \right\} \right]\\ \nonumber
	&-MD_{pp}(t)\left[x,\left[x,\rho(t) \right]\right]-D_{xp}(t)\left[x,\left[p,\rho(t) \right]\right],
\end{align}
with a time-dependent Hamiltonian $H_R$ and time-dependent coefficients $ D_{xx}$, $D_{pp}$, and $D_{x,p}$ \cite{Hu1992,OConnell2010,FlemingHu}. 
Note the absence of a memory kernel when written in this form,
which is implicit in the coefficients.}

\je{There are several closely related meaningful ways to quantify Markovianity of a process \cite{Markovian1,Markovian2,Markovian3}, all essentially
deriving from infinite divisibility of the dynamical map (physically originating from short bath correlation times). In precisely this spirit, we capture non-Markovianity by the extent to which 
the  right hand side of Eq.\ (\ref{HPZequation}) deviates from a valid Lindblad generator (a rigorous treatment is presented \jens{in note 3}
in the supplementary material). The measure taken is
\begin{eqnarray}
	\xi:&=& \min\left\{0, \lim_{t\rightarrow\infty} 
	\frac{- \lambda_{\rm min} (\Xi(t))}{\| \Xi(t)\|}
	\right\},\\
	\Xi(t)&=&\left(\begin{array}{cc}
	2MD_{pp} (t)& D_{xp}(t)+i\gamma(t)\\
	D_{xp}(t)-i\gamma (t)& 0
\end{array}\right).
\end{eqnarray}
For an Ohmic spectral density with high frequency cutoff we find that $D_{xp}(\infty)$ is very close to zero;
in fact $\xi$ is of the order of $10^{-15}$ for all other parameters chosen as in the experiment.
However, our result for the slope at $I(\Omega)$ gives a lower bound of $\xi> 1.1\times 10^{-6}$. 
\je{This shows that the dynamics sharply deviates from a Markovian
one}.}
In other words, our analysis unambiguously
shows that the heat bath of the micro-mechanical oscillator is not consistent with Markovian damping of a quantum harmonic oscillator in the high temperature limit.

\begin{figure}[t]
\medskip
\centering
\includegraphics[width=0.95\linewidth]{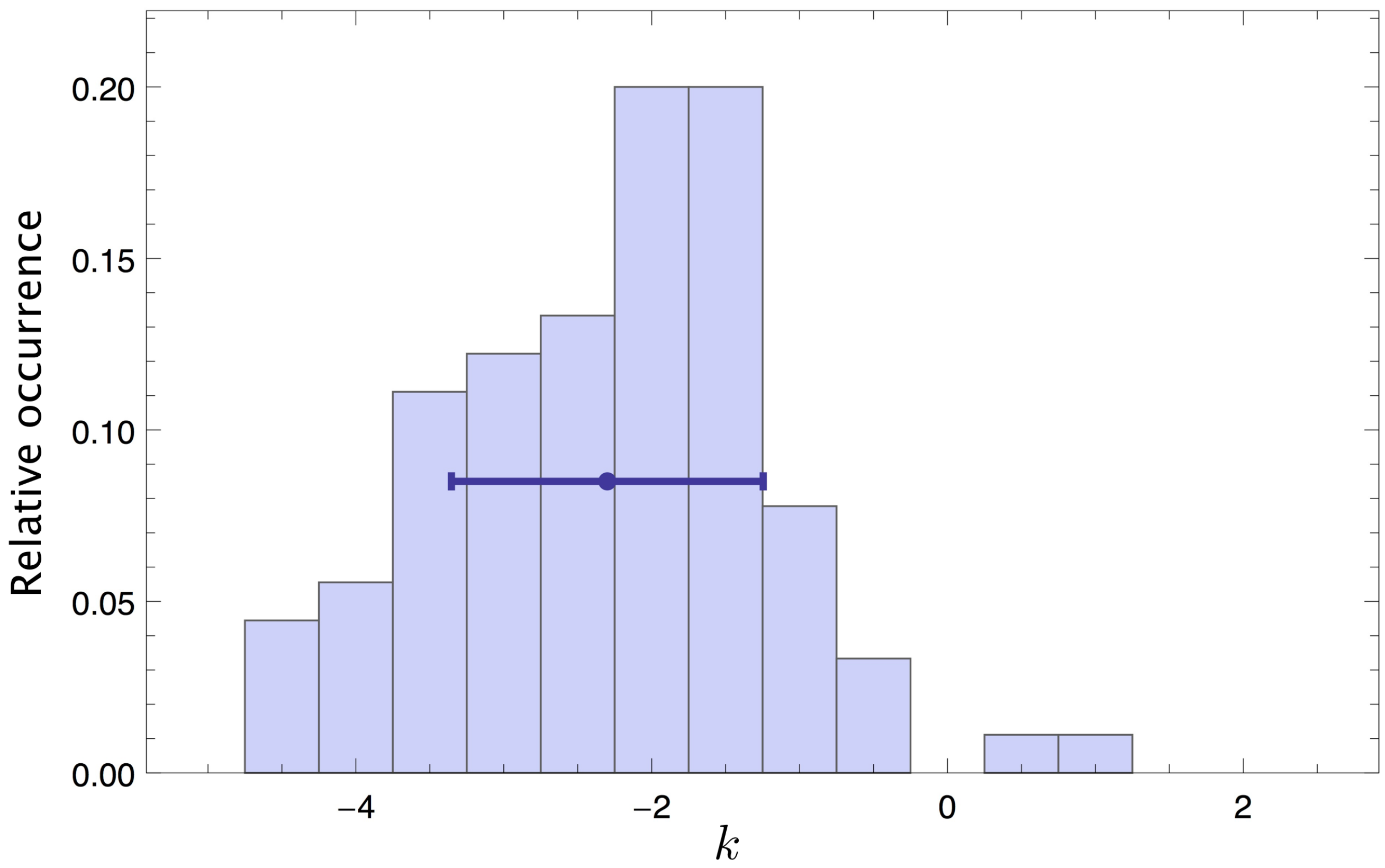}
\caption{\textbf{Estimated coefficients.} Depicted is the histogram of best estimated coefficients $k$ in the local approximation 
\je{within $[\omega_{\rm min},\omega_{\rm max}]$} of the spectral density by $I(\omega)=C \omega^k$, showing a statistically significant deviation from the Ohmic situation of $k=1$.}
\label{Fig:histogram}
\end{figure}

\section{Discussion}

While we do not expect effects of finite-dimensional bath components resulting, e.g., from two-level fluctuators, to measurably influence the result~\cite{Verhagen2012,Ustinov,Schlosshauer2008}, we cannot rigorously exclude such contributions. We can yet strictly and unambiguously falsify the common assumption of a harmonic Ohmic heat bath. 
Our specific situation is rather described by highly sub-Ohmic damping. 
We strongly emphasise that our analysis does not rely on any assumption about the resonator geometry.
We may, however, still speculate as to why this strongly sub-Ohmic damping is being found. It seems plausible
that the specific geometry of the slab used contributes to this non-orthodox decoherence. Indeed, sub-Ohmic 
spectral densities have been computed in a phononic mode analysis of low-dimensional slabs \cite{Wilson-Rae2008b}.
However, we also expect intrinsic decoherence mechanisms
to be relevant.

It is known that in non-Ohmic baths the coefficients of the master equation governing the dynamics are becoming strongly time-dependent \cite{Hu1992}. 
 This means that while the steady-state properties of a mechanical system may be modified only in a mild way --  the deviations 
of the measured spectrum from Eq.\ (\ref{Spectrum}) for Ohmic spectral densities are small -- one should expect larger deviations for predictions in 
time-dependent situations \cite{Mari2009}. It has been pointed out recently that such non-Markovian quantum noise can significantly influence the ability to generate quantum entanglement~\cite{Ludwig}. Indeed, intricate memory effects come into play in case of non-Markovian dynamics, giving rise to a picture of  decoherence beyond basic rate equations.

Finally, our findings complement related research in mechanical engineering. 
It is known that damping due to internal materials losses can be vastly different from a purely velocity dependent damping term as typically assumed for a simple harmonic oscillator. Specific models for such non-viscous damping, a prominent model being that of `structural' or frequency-independent damping \cite{Saulson1990}, have been extensively studied in the context of both gravitational wave detection~\cite{Saulson1990,Gonzalez1995,Kajima1999} and measurements of the gravitational constant~\cite{Bernardini1999,Yang2009}, where thermal noise in the DC tail of a mechanical resonance poses limits on the achievable sensitivity. In turn, while the accurate measurement of internal friction and the analysis of their origin remains a challenging task, broadband thermal noise measurements have become an important input for the design and engineering of high-$Q$ micro- and nano-mechanical resonators~\cite{Sosale2012}. 
This is also important for macroscopic systems such as end mirrors for optical reference cavities or gravitational wave detectors.

Our approach adds two new aspects: First, our analysis provides a direct link to `Markovianity'~\cite{Markovian1,Markovian3,Markovian2} as a statistical property of the environment of a quantum harmonical oscillator. Second, we exploit the enhancement of the thermal noise in the vicinity of the mechanical resonance, instead of probing thermal noise over a broad frequency band. This provides a local estimate of the thermal bath characteristics, which is the relevant property for non-Brownian dynamics. In a next step, combining this method with a sweep in resonance frequency~\cite{Jockel2011} could provide direct, full broadband mechanical spectroscopy of the thermal bath spectral density, in a `tomographic approach'. Our system identification approach is also model-independent, i.e., we do not make any prior assumptions on the underlying nature of the dissipation or on the specific shape of the thermal noise spectral density (other than assuming harmonicity). Although the current study is performed at room temperature, in the `classical' regime, it can be directly applied to other mechanical resonators that operate close to or in the quantum regime~\cite{Rocheleau2010,Teufel2011b,Safavi-Naeini2012,Teufel}.

In summary, we have introduced a versatile method to directly probe the spectral density of the heat bath of a micro-mechanical resonator. 
We demonstrate that the common assumption of Markovian Brownian motion does not hold. This opens the way towards systematic studies of individual dissipation channels such as two-level fluctuators~\cite{Verhagen2012,Ustinov}. In combination with the possibility to geometrically modify the phonon spectrum~\cite{Wilson-Rae2008b,Anetsberger2008,Cole2011,CrossLifshitz,Eva} this would allow for full reservoir engineering of quantum harmonic oscillators. We hope that the present work stimulates such further experimental analysis of unorthodox decoherence phenomena opening up alongside technological development.

\subsection*{Acknowledgements}

We acknowledge discussions with 
C.\ Fleming,
B.-L.\ Hu,
D.\ Loss,
A.\ Mari, 
G.\ Milburn,
J.-P.\ Paz,
P.\ Rabl, 
I.\ Wilson-Rae, and
A.\ Xuereb,
and contributions to the statistical estimation by K.\ Kieling. This work has been supported by the EU (SIQS, MINOS, ITNcQOM, AQuS, AQuS, IQOEMS,
RAQUEL and a Marie Curie fellowship of S.~G.), the EURYI, the 
ERC (StG of M.~A.\ and J.~E.), the BMBF (QuOReP), the Austrian Science Fund FWF (START, FOQUS), and the Alexander von Humboldt Foundation (Bessel Research Award of M.~A.).

\subsection*{Authors' contributions}

\jens{J.\ Eisert and M.\ Aspelmeyer have planned this research, J.\ Eisert and A.\ Trubarov have performed
the theoretical analysis and have performed the relevant research on non-Markovian quantum Brownian motion,
S.\ Gr{\"o}blacher, G.\ D.\ Cole, and M.\ Aspelmeyer have performed the experiment, N.\ Prigge and J.\ Eisert
have performed the data analysis and have performed the research on quantifying non-Markovianity, 
with help of the experimental team. All authors have contributed to writing the manuscript. }



\section{Supplementary material}

\subsection*{\jens{Supplementary note 1:\\ Noise-noise correlations for the damped mode}}

{\footnotesize

\je{In this work, we allow for and discuss an arbitrary harmonic model for arbitrary spectral densities of the bath, where special emphasis is put on the weak 
coupling and high temperature limit, but no further assumptions are being made.}
We start from the Hamiltonian of the Caldeira-Leggett-model (or Ullersma model)
for quantum Brownian motion~\cite{Caldeira1983,Hu1992,NewPaz}, 
so the standard quantum mechanical damped oscillator,
\begin{eqnarray}\label{Mechanical}
	H_{m} &=& \frac{p^2}{2m}+ \frac{1}{2}m \Omega^2 q^2 + \sum_n 
	\left(
	\frac{p_n^2}{2m_n} + \frac{1}{2} m_n \omega_n^2 q_n^2
	\right)\nonumber\\
	&+& q \sum_n c_n q_n.
\end{eqnarray}
At this point no assumption is made with respect to the coupling except from it being linear.
The equations of motion of the canonical coordinate of the \je{distinguished} oscillator are given by
\begin{eqnarray}\label{EOM}
	\ddot q (t) + \Omega^2 q(t)+ \frac{2}{m} \int_0^t ds \eta(t-s) q(s) = \frac{f(t)}{m}.
\end{eqnarray}
\je{In this equation}, the inhomogeneity is given by
\begin{eqnarray}
	f(t) &=& -\sum_n c_n \left(
	q_n(0) \cos(\omega_n t) + \frac{p_n(0)}{m_n}\frac{\sin(\omega_n t)}{\omega_n}
	\right),
\end{eqnarray}
whereas the damping kernel $\eta$ is
\begin{eqnarray}
	\eta(s) &=&\frac{d}{ds}\nu(s),\\
	\nu(s) &=& \int_0^\infty d\omega \frac{I(\omega)}{\omega} \cos(\omega s),
\end{eqnarray}
in terms of the spectral density
\begin{eqnarray}
	I(\omega) &=& \sum_n \delta(\omega-\omega_n) \frac{c_n^2}{2m_n \omega_n}.
\end{eqnarray}
In the Ohmic regime, so for a spectral density linear in $\omega$ until a finite \je{but large} cut-off frequency $\Lambda>0$, 
this damping kernel is for large $\Lambda$ arbitrarily well approximated by an expression of the form
\begin{equation}
	\eta(t) = \gamma(\infty) m\delta'(t), 
\end{equation}
so Eq.~(\ref{EOM}) takes a form local in time. Returning to the general case, the exact two-point correlation functions of the thermal force are found to be
\begin{eqnarray}
	\langle
	f(t) f(s)
	\rangle& =& \int_{-\infty}^\infty e^{i\omega(t-s) } \frac{\hbar}{2} I(\omega) 
	\nonumber\\
	&\times &\left(
	\coth\left(\frac{\hbar \omega}{2k_B T})\right) -1
	\right)d\omega,
\end{eqnarray}
for $t,s\geq 0$, where for simplicity of notation we have extended the definition of the spectral density to $I:\rr\rightarrow \rr^+$ by taking
$I(-\omega)=-I(\omega)$. 

This expression, valid in general without any approximation, can be cast into a time-local form, albeit the dynamics being non-Markovian. 
For this, one has to formulate the Green's function $G:\rr\rightarrow \rr$ of the problem. One arrives at a differential equation of the form
\begin{eqnarray}
	\ddot q (t) + \Omega^2(t)  q(t)+  \gamma(t) \dot q(t) = \frac{\bar f(t)}{m}
\end{eqnarray}
for  $t\geq 0$, where now the time-dependent damping is found to be
\begin{eqnarray}
	\gamma(t) &=& \frac{G(t) \dddot G(t) - \dot G(t) \ddot G(t)}{\dot G(t)^2 - G(t) \ddot G(t)},
\end{eqnarray}
and
\begin{eqnarray}
	\Omega^2 (t) = \frac{\ddot G(t)^2 -  \dot G (t) \dddot G(t)}{\dot G(t)^2- G(t) \ddot G(t)}.
\end{eqnarray}
The inhomogeneity becomes
\begin{equation}
	\bar f(t) = \left(\partial_t^2 + \gamma(t) \partial_t + \Omega(t)^2 \right)\int_0^t G(t-s) f(s) ds.
\end{equation}
It can now be shown, and is physically plausible, that these quantities take their asymptotic values for large times,
\begin{equation}
	\lim_{t\rightarrow\infty} \gamma(t)= \gamma(\infty), 
	\lim_{t\rightarrow\infty}  \Omega(t) = \Omega(\infty).
\end{equation}
These expressions are exact and no approximations have been made until this point.
For large times $t$, the expression
\begin{eqnarray}
	\ddot q (t) + \Omega^2(\infty)  q(t)+  \gamma(\infty) \dot q(t) 
\end{eqnarray}
is arbitrarily well approximated by 
${\bar f(t)}/{m}$.
In fact, the correlation function of this modified driving can readily be found by going into the Fourier domain, with convention taken
\begin{equation}
 	\tilde F(\omega)= \int_{-\infty}^\infty
	F(t) e^{-i\omega t} dt.
\end{equation}
An evaluation of this expression reveals that the Fourier transform of the Green's function is given by
\begin{equation}
\tilde G(\omega) = \frac{1}{-\omega^2 + 2 \hat \eta(\omega)/m + \Omega^2},
\end{equation}
where we have defined $\hat{\eta}$ and analogously $\hat{\nu}$ as
\begin{equation}
	\hat{\eta}(\omega)=\int_0 ^{\infty} dt  \eta (t) e^{-i\omega t}.
\end{equation}
The expected two-time correlation function of $\bar f$ is then computed to be
\begin{eqnarray}\label{cc}
	\langle
	\bar f(t) \bar f(s)
	\rangle& =& \int_{-\infty}^\infty 
	e^{i\omega(t-s) } \frac{\hbar}{2} I(\omega)\nonumber \left(
	\coth\left(\frac{\hbar \omega}{2k_B T}\right) -1
	\right) \\
	&\times &((\Omega(\infty)^2-\omega^2)^2 + \gamma(\infty)\omega^2)|\tilde{G}	(\omega)|^2 d\omega
	\\ &=&\int_{-\infty}^\infty 
	e^{i\omega(t-s) } \frac{\hbar}{2} I(\omega) 
	\nonumber\\
	&\times &
	\frac{(\Omega(\infty)^2-\omega^2)^2 + \gamma(\infty)\omega^2}{
	(\Omega^2-\omega^2+ 2\text{re}(\hat\eta(\omega))/m)^2
	+ (2\text{im}(\hat\eta(\omega))/m)^2}\nonumber\\
	&\times &\left(
	\coth\left(\frac{\hbar \omega}{2k_B T}\right) -1
	\right)d\omega,
\end{eqnarray}
for $t,s\geq 0$. We also find in terms of $\nu$, rather than in $\eta$,
\begin{eqnarray}\label{ne}
	\langle \bar f(t) \bar f(s)\rangle &=& \int_{-\infty}^\infty 
	\frac{
	(\Omega(\infty)^2-\omega^2)^2+ \gamma^2(\infty)\omega^2
	}{\left(K^2 -\omega^2 + 2 \omega 
	 {\rm im}(\hat \nu(\omega))/m
	\right)^2 + 
	\left(2 \omega
	 {\rm re}(\hat \nu(\omega))
	 /m
	\right)^2
	}
	 \nonumber\\
	&\times & \frac{\hbar}{2}I(\omega)
	\left(
	\coth\left(
	\frac{\hbar \omega\beta}{2}
	\right)
	-1
	\right) e^{i\omega(t-s)} d\omega,
\end{eqnarray}
where 
\begin{equation}
	K^2 = -\frac{2}{m}
	\nu(0) + 
	\Omega^2,
\end{equation}
and $\beta=1/(k_BT)$.
This is still an exact expression. The real part is now identified to be
\begin{equation}
	 {\rm re}(\hat \nu(\omega))= \frac{\pi I(\omega)}{2\omega}.
\end{equation}
The weak coupling approximation amounts to approximating
\begin{equation}
	\Omega(\infty)^2 \approx K^2,  
	\gamma(\infty) \approx \frac{\pi I(K)}{m K},
\end{equation}
which is true if
\begin{eqnarray}
	\left|
	\frac{2}{m}
	\partial_\omega \hat \nu(\omega) |_{\omega=K}
	\right| \ll 1,\\
	\left|
	\frac{1}{m}
	\hat \nu(K) 
	\right| \ll K
\end{eqnarray}
and if the imaginary part is negligible. This is the standard weak coupling limit \cite{FlemingHu}, 
which is the only approximation made in the discussion of quantum Brownian motion.
Note that this weak coupling limit does not require the coupling to be so weak for the rotating wave approximation or the `quantum optical limit' (see Subsection \ref{2M}) to be valid.
With these approximations, one finds that 
\begin{equation}
	\frac{
	(\Omega(\infty)^2-\omega^2)^2+ \gamma^2(\infty)\omega^2
	}{\left(K^2 -\omega^2 + \omega 
	\frac{2}{m} {\rm im}(\hat \nu(\omega))
	\right)^2 + 
	\left(\omega
	\frac{2}{m}  {\rm re}(\hat \nu(\omega))
	\right)^2
	}
\end{equation}
is extraordinarily well approximated by unity, meaning that the contribution of the spectral density to the two-time correlation function of $\bar{f}$ originates 
only from the nominator. This is basically the reason, why in Eq.\ (\ref{ne})
$I(\omega)$ appears only in the numerator. Within this approximation, we have that
$\langle \bar f(t)^2\rangle \approx \langle f(t)^2\rangle$ to a very good approximation.

\subsection*{\jens{Supplementary note 2:\\ Relating the spectra of the mirror and the output light}}

We now discuss the situation where the harmonic oscillator described above is coupled to a laser field within an optical cavity. The vibrational mode of a high reflective micro-mirror is modelled as a damped harmonic oscillator as in Eq.~(\ref{Mechanical}). The micro-mirror together with another solid mirror forms an optical cavity, which is driven by a laser beam. The total Hamiltonian of a harmonic oscillator coupled to thermal bath and laser field within a driven optical cavity is given by
\begin{equation}
H = H_m +\hbar \omega_c a^\dagger a - \hbar g_0 a^\dagger a q +i\hbar E(a^\dagger e^{-i\omega_0 t}- ae^{i\omega_0 t}),
\end{equation}
where $a$ is the annihilation operator of the optical mode, $g_0 = \omega_c/L$ is the coupling constant of the mechanical to the optical mode. $\omega_c$ is the resonance frequency of the cavity with length $L$ and decay rate $\kappa$. $H_m$ is the complete Hamiltonian of the distinguished mechanical mode with its environment as in Eq.~(\ref{Mechanical}). 
\begin{equation}
|E| = \sqrt{2W\kappa/(\hbar \Omega)},
\end{equation}	
where $W$ is the input power of the laser with frequency $\omega_0$. The Heisenberg picture equations of motion formulated in the interaction picture 
with respect to $\hbar \omega_0 
a^\dagger a$ become, suppressing time dependence,
\begin{eqnarray}
	\dot q  & = & \frac{p}{m},\\
	\dot p  & = & -m \Omega^2 q - \sum_n c_n q_n + \hbar g_0 a^\dagger a,\\
	\dot a  & = & -(\kappa + i \Delta_0) a + i g_0 a q + E + \sqrt{2\kappa }a^{\rm in},\\
	\dot q_n & = &\frac{p_n}{m_n},\\
	\dot p_n & = &- m_n \omega_n^2 q_n - c_n q,
\end{eqnarray}
where $\Delta_0= \omega_c - \omega_0$. In the weak-coupling limit of the previous subsection, the equations of motion turn into
\begin{eqnarray}
	\ddot q  + \gamma(\infty) \dot q + \Omega(\infty)^2 q & = & \frac{\bar f}{m} + \frac{\hbar g_0}{m}a^\dagger a,\\
	\dot a  & = & -(\kappa + i \Delta_0) a + i g_0 a q + E + \sqrt{2\kappa }a^{\rm in},
\end{eqnarray}
again suppressing time dependence. Based on these expressions, one can proceed exactly as presented in Ref.\ \cite{Genes},
with the Ohmic bath being replaced by this general thermal bath. In order to progress, it is helpful
to make use of dimensionless quantities,
\begin{eqnarray}
	Q = \frac{q}{l}, 
	P = \frac{pl}{\hbar}, 
	l= \sqrt{\hbar/(m\Omega(\infty))}
\end{eqnarray}
and to define 
	$G_0 = g_0 l$.
Following Ref.\ \cite{Genes},
one arrives at an expression which is for large times $t,s$ well approximated by
\begin{equation}
	\langle \delta Q(t)\delta Q(s)\rangle \approx \int_{-\infty}^\infty \frac{d\omega}{2\pi}
	|\chi_{\rm eff}^\Delta(\omega)|^2
	\left(
	S_{\rm th} (\omega) +
	S_{\rm rp} (\omega,\Delta) 
	\right)e^{i\omega(t-s)},
\end{equation}
for the deviations from the asymptotic steady state value
\begin{equation}
	\delta Q = Q - \lim_{t\rightarrow\infty} \langle Q\rangle = Q - 
	\frac{G_0 |\alpha_s|^2}{\Omega(\infty)}
	,
\end{equation}
where
\begin{eqnarray}
	S_{\rm th}(\omega) &=&\frac{\pi I(\omega)}{m \Omega(\infty)}\left(\coth\left(
	\frac{\hbar \omega}{2 k_B T}
	\right)-1
	\right),\\
	S_{\rm rp}(\omega,\Delta)&=& \frac{2 \kappa  G_0^2|\alpha_s|^2}{\Delta^2+ \kappa^2+ \omega^2+ 2\Delta\omega},\\
	\chi_{\rm eff}^\Delta(\omega)&=&\Omega(\infty)\left(
	\Omega(\infty)^2 + i \gamma(\infty) \omega - \omega^2 - \frac{2G_0^2|\alpha_s|^2 \Delta\Omega(\infty)}{\Delta^2 + (\kappa + i\omega)^2}
	\right)^{-1}.
\end{eqnarray}
Here, $\Delta$ and $\alpha_s$ are implicitly defined as
\begin{equation}
	\Delta= \Delta_0 - \frac{G_0^2|\alpha_s|^2}{\Omega(\infty)},  
	\alpha_s = \lim_{t\rightarrow\infty} \langle a\rangle = \frac{E}{\kappa + i\Delta}.
\end{equation}
This is different from the expression in Ref.~\cite{Genes} in that both the thermal noise spectrum $S_{\rm th}$ as well as the radiation pressure noise spectrum $S_{\rm rp}$ are being altered.

The final step is to include the actual measurement of the opto-mechanical system and how the information about the motion of the mechanics can be obtained from the measurement of the light leaking out of the cavity. As before, this entire apparatus is assumed to be well-characterised and known, which is a very reasonable assumption for the present experiment.

In the experiment the light leaking out of the optical cavity, referred to as `signal', is measured by homodyne detection. This means that signal is mixed on a 50:50 beam-splitter with a second, much stronger laser beam. This second beam has the same frequency as the light driving the cavity and is referred to as the `local oscillator'. The intensities measured in both arms are then electronically subtracted. The result
is a voltage, which apart from delta-correlated measurement noise is proportional to some general quadrature of the signal, depending on the phase between the signal and the local oscillator and is, apart from noise, proportional to the corresponding intra-cavity quadrature. In the the parameter regime relevant here, it turns out that the output quadrature $\delta Y^{\rm out}$ is proportional to $\delta Q$ subjected to additional noise. Hence, by measuring this quadrature of the signal, information about the mechanical motion and therefore about the thermal
bath driving the mechanics can be extracted.

Again following Ref.~\cite{Genes}, one arrives in the regime that is experimentally relevant at the expression for the spectrum of the output light measured with quantum efficiency $\zeta>0$ that is very well approximated by the expression in dimensionless units
\begin{equation}
	S_{\delta Y^{\rm out}}(\omega) \approx \frac{8k_B T \pi\zeta G_0^2 |\alpha_s|^2 \Omega(\infty)}{m\hbar \kappa}
	\frac{I(\omega)}{\omega\left(
	(\Omega(\infty)^2-\omega^2)^2 + (\gamma(\infty)\omega)^2
	\right)}.
\end{equation}
That is to say, by detecting the output light, one can immediately obtain information on the spectral density $I$ of the unknown decohering environment. 

\subsection*{\jens{Supplementary note 3:\\ Non-Markovian dynamics}}\label{2MNew}

\je{For what follows, we will put properties of spectral densities into contact with 
non-Markovian features of the resulting dynamics. 
In order to make this link precise, we will first discuss how Markovian and non-Markovian dynamics 
can be captured for harmonic systems. 
Generally speaking, there are several closely related meaningful ways to quantify the non-Markovianty of a process 
\cite{Markovian1,Markovian2,Markovian3}, all essentially deriving from infinite divisibility of the dynamical map, the latter being 
defined as
\begin{equation}
	\rho \mapsto T_t(\rho)= \rho(t)
\end{equation}
for states $\rho$. The mathematical property of infinite divisibility can be interpreted in physical terms:
Markovian dynamics is forgetful dynamics, one that results from an interaction with a heat bath that does not keep a memory, 
a property that in turn is originating from short bath correlation times. Fully Markovian dynamics is always an abstraction, even though
often, dynamics can be Markovian to an extraordinarily good approximation. By virtue of Lindblad's theorem \cite{Lindblad,Kossakowski},
Markovian dynamics is reflected by a time evolution governed by a master equation
\begin{equation}
	\frac{d}{dt}\rho(t) = {\cal L}(\rho(t)),
\end{equation}
with a time-independent generator of so-called Lindblad form, 
\begin{equation}
	{\cal L}(\rho) =\sum_j\left(
	L_j \rho L_j^\dagger - \frac{1}{2}\{L_j^\dagger L_j,\rho\}
	\right).
\end{equation}
Note that in some recent treatments of Markovianity, even time dependent equations of motion are considered Markovian -- since we are interested in the long-time limit here, 
this distinction is of no relevance for our purposes, however. In the focus of attention are harmonic systems of a single mode, coupled to a harmonic bath \cite{GaussianMarkovian}.}

\je{Starting point of the analysis is the exact master equation for quantum Brownian motion for an arbitrary spectral density in a general environment \cite{Hu1992},
given by
\begin{align*}
\frac{d}{d t}\rho(t)=&-i\left[H_{R}(t),\rho(t) \right]-i\gamma (t)\left[x,\left\{ p,\rho(t) \right\} \right]\\
&-MD_{pp}(t)\left[x,\left[x,\rho(t) \right]\right]-D_{xp}(t)\left[x,\left[p,\rho(t) \right]\right] .
\end{align*}
The time-\je{dependent} coefficients are completely determined by the spectral density, however in \je{general in} an extraordinarily 
complicated way. Note that for harmonic systems,
the non-Markovian character and the memory are implicitly
entirely incorporated by the time dependence of the coefficients of the 
master equation, and no further memory kernel is necessary to keep generality.
As this equation is quadratic in the canonical coordinates, one can easily deduce the dynamical law for the $2\times 2$ real 
covariance matrix $\Gamma$ 
with time dependent entries
\begin{equation}
	\Gamma (t) = \left(
	\begin{array}{cc}
	2 \langle x^2 \rangle (t)& \langle x p +p x\rangle(t)\\
	 \langle xp +px\rangle(t) & 2  \langle p^2\rangle(t) 
	\end{array}
	\right) ,
\end{equation}
as 
\begin{align}
\label{gammaHPZ}
	\frac{d}{dt}\Gamma(t)&=- h(t)\Gamma(t)-\Gamma(t)h^T(t)+D(t),
	\\
	\nonumber  h(t)&=\left(\begin{array}{cc}
	0 & -\frac{1}{M}\\
	M\Omega_{r}^{2}(t) & 2\gamma(t)
\end{array}\right),\\
	\quad D(t)&=\left(\begin{array}{cc}
	0 & -D_{xp}(t) \\
	-D_{xp} (t) & 2MD_{pp} (t)
\end{array}\right).
\end{align}
\je{One can characterise the entire interaction with the environment by a Hermitian $2\times 2$ matrix $\Gamma$} 
that one can deduce from this differential equation. For this, we represent $H_R(t)$ as
the quadratic form
\begin{equation}
	H_R(t)=\left(\begin{array}{cc}
	x & p\end{array}\right)h_R(t)\left(\begin{array}{c}
	x\\
	p
	\end{array}\right) 
\end{equation}
and define the symplectic matrix as
\begin{equation}
	\sigma =\left(\begin{array}{cc}
	0 & 1\\
	-1 & 0
	\end{array}\right).
\end{equation}	
It then follows that
\begin{eqnarray}
	h^T(t) &=&\left(2h_R(t) -\text{im}(\Xi(t) )\right)\sigma ,\\
	 D(t) &=&\sigma^T \text{re}(\Xi(t))\sigma.
\end{eqnarray} 
We find that the system-environment interaction is then specified by 
\begin{equation}
	\Xi(t)=\left(\begin{array}{cc}
	2MD_{pp} (t)& D_{xp}(t)+i\gamma(t)\\
	D_{xp}(t)-i\gamma (t)& 0
\end{array}\right).
\end{equation}
The long-time dynamics is Markovian if and only if $\Xi(\infty)$ corresponds to a master equation in Lindblad form, which
in this case is equivalent with $\Xi(\infty)$ being positive semi-definite. Hence, a meaningful measure of non-Markovianity is -- in case of long-time
non-Markovian dynamics -- 
the absolute value of the smallest eigenvalue of $\Xi(\infty)$, normalised by the operator norm of $\Xi(\infty)$. More precisely,
the measure used is 
\begin{equation}
	\xi= \min\left\{0, \lim_{t\rightarrow\infty} 
	\frac{- \lambda_{\rm min} (\Xi(t))}{\| \Xi(t)\|}
	\right\}.
\end{equation}
This quantity, which is particularly transparent in this case of harmonic dynamics,
 can easily be related to other measures of non-Markovianity discussed in the literature 
\cite{Markovian1,Markovian3,Markovian2,GaussianMarkovian}. 
 If $\xi =0$, the dynamics is Markovian in the long-time limit. Otherwise, it is non-Markovian, and more significantly so the larger this quantity.
As $\lim_{t\rightarrow \infty}D_{pp}(t)>0$, one finds the simple expression
\begin{equation}\label{xi}
	\xi= \frac{(1+\mu )^{1/2}-1}{ (1+ \mu)^{1/2}+1},
\end{equation}
with
\begin{equation}
\label{measure}
	\mu = \lim_{t\rightarrow\infty} \frac{D_{xp}^{2}(t)+\gamma^{2}(t)}{M^{2}D_{pp}^{2}(t) }.
\end{equation}
Since $\xi$ is a simple monotonous function of $\mu$, one can as well quantify non-Markovianity of the long-time dynamics in terms of $\mu$. }

\subsection*{\jens{Supplementary note 4:\\ Relating non-Markovianity to spectral densities}}\label{2M}

\je{In this subsection, we relate notions of non-Markovianity of dynamics with properties of spectral densities. This link has already been rather well established \cite{Hu1992,OConnell2010},
in that it is well known that the high-temperature Ohmic setting corresponds to the Markovian limit to a very good approximation.
 Here, in order
to further strengthen the claim of the main text, we significantly extend this link and make it more quantitative and precise: We show how a deviation from this behaviour
can be directly and quantitatively related to non-Markovian features. That is to say,
we make the link between Ohmic spectral densities and Markovian dynamics in the above precise sense quantitative.}

\je{
Ref.\ \cite{Hu1992} presents the 
following simplified expressions for the coefficients in the weak-coupling limit
\begin{align}
	\delta\Omega^{2}\left(t\right)&=2\int_{0}^{t}ds \eta(s)\cos\left(\Omega s\right),\,\,\,
	\Omega^2(t) =\Omega^2+ \delta\Omega^2(t), 
		\\
	\gamma\left(t\right)&=-\frac{1}{\Omega}\int_{0}^{t}ds \eta(s)\sin\left(\Omega s\right),\\
	D_{xp}\left(t\right)&=\frac{1}{\Omega}\int_{0}^{t}dsn(s)\sin\left(\Omega s\right),\\
	m D_{pp}\left(t\right)&=\frac{1}{\Omega}\int_{0}^{t}ds n(s)\cos\left(\Omega s\right).
\end{align}
The noise and damping kernel (in a notation adapted to the present work)
are given by
\begin{align}
	n(s)&=\int_{0}^{\infty}d\omega  I(\omega)\coth\left(\frac{\hbar\omega}{2 k_B T}\right)\cos\left(\omega s\right) ,\\
	\eta(s)&=\frac{d}{ds}\left(\int_{0}^{\infty}d\omega \frac{ I(\omega)}{\omega}\cos\left(\omega s\right)\right)=-\int_{0}^{\infty}d\omega   I(\omega)\sin\left(\omega s\right) .
\end{align}
}

\je{So far, we have merely recapitulated properties of general quantum Brownian motion in the weak coupling limit. We now turn to making the link of a
deviation from the high-temperature Ohmic setting to non-Markovian dynamics precise. In order to do so in a most transparent fashion, we first state the mild
and natural assumptions made on the spectral density $I:\rr^+\rightarrow \rr^+$.
We make the following assumptions:
\begin{itemize}
\item[(i)]The function 
\begin{equation}
	\omega \mapsto I(\omega) \coth\left(\frac{\hbar\omega}{2 k_B T}\right) 
\end{equation}
is in $L^1\left([0,\infty )\right)$.
\item[(ii)] The spectral density can be approximated in a vicinity of $\Omega$ by an affine function (or a power law), i.e., it is not rapidly oscillating.
\end{itemize}
The first assumption is necessary in order to get a well-defined noise kernel, 
such that the occurring integrals are convergent. 
The second assumption is required for the statistical analysis. In addition to these assumptions on the spectral density, we will invoke the above weak coupling 
and a high temperature
approximation. Both the weak coupling as well as the high temperature approximation are valid to overwhelming accuracy for the experiment at hand.}
\je{A standard calculation  shows that
\begin{eqnarray}		
	\underset{t\rightarrow \infty}{\lim}\gamma (t) &=& \gamma(\infty) = \frac{\pi}{2\Omega}I(\Omega ),\\
	\underset{t\rightarrow \infty}{\lim}m D_{pp}(t)&=&m D_{pp}(\infty)= \frac{\pi}{2\Omega}I\left(\Omega\right)
	\coth\left(\frac{\hbar\Omega}{2 k_B T}\right).
\end{eqnarray}
The computation of $ \underset{t\rightarrow \infty}{\lim}D_{xp}(t)$ is, however, more subtle and we will make use of assumption
(ii). We find
\begin{equation}
D_{xp}(t)=\frac{1}{\Omega}\int_{0}^{t}ds\left(\int_{0}^{\infty}d\omega  I(\omega)\coth\left(\frac{\hbar\omega}{2 k_B T}\right)\cos\left(\omega s\right)\right)\sin\left(\Omega s\right).
\end{equation}
By assumption (i), we may use Fubini's theorem to get
\begin{eqnarray}
	D_{xp}(t)&=&\frac{1}{\Omega}\int_{0}^{\infty}d\omega  I(\omega)\coth\left(\frac{\hbar\omega}{2 k_B T}\right)\\
	&\times&
	\frac{1}{2}\left\{ -\frac{\cos\left(\left(\Omega+\omega\right)s\right)}{\Omega+\omega}\big|_{0}^{t}-\frac{\cos\left(\left(\Omega-\omega\right)s\right)}{\Omega-\omega}\big|_{0}^{t}\right\} \nonumber\\
	&=&\frac{1}{2 \Omega}\int_{0}^{\infty}d\omega   I(\omega)\coth\left(\frac{\hbar\omega}{2 k_B T}\right)  \bigg(\frac{1}{\Omega+\omega} \nonumber\\
	& +& \frac{1}{\Omega-\omega}\left(1-\cos\left(\left(\Omega-\omega\right)t\right)\right)-\frac{1}{\Omega+\omega}\cos\left(\left(\Omega+\omega\right)t\right)\bigg).\nonumber
\end{eqnarray}
If 
\begin{equation}
	\omega\mapsto I(\omega) \coth\left(\frac{\hbar\omega}{2 k_B T}\right)
\end{equation}
is in $L^1\left([0,\infty)\right)$, by assumption (i), 
then so is 
\begin{equation}
	I(\omega)  \coth\left(\frac{\hbar\omega}{2 k_B T}\right) \frac{1}{\omega+\Omega}. 
\end{equation}
Therefore, 
for $t\rightarrow \infty$ the last term vanishes by the Riemann-Lebesgue Lemma. We cannot proceed similarly for 
the other oscillatory part because 
$1/x$ is not integrable over $\rr ^+$. However, as $\lim_{x\rightarrow \infty}{(1-\cos(x))}/{x}=0$ the integral is convergent and we can investigate the influence of the local behaviour of the spectral density around the resonance frequency.
}
\je{We proceed by splitting the domain of integration into a part 
\begin{equation}
	U_{\Omega}=(\Omega-\delta,\Omega+\delta) 
\end{equation}
close to resonance and its complement for a small $\delta>0$. 
We can then again make use of the Riemann-Lebesgue Lemma for the oscillatory part over $\mathbb{R}^+\setminus \left(\Omega -\delta , \Omega + \delta \right)=:U_{\Omega}^c$,
to obtain
\begin{eqnarray}
D_{xp}^{\rm res}(t)&=&\frac{1}{2\Omega}\int_{\Omega -\delta}^{\Omega + \delta}d\omega   I(\omega)\coth\left(\frac{\hbar\omega}{2 k_B T}\right) \\
&\times  &\left\{ \frac{1}{\Omega+\omega}+\frac{1}{\Omega-\omega}\left(1-\cos\left(\left(\Omega-\omega\right)t\right)\right)\right\} ,\\
D_{xp}^{\rm off}(\infty)&=&\int _{U_{\Omega}^c}d\omega  I(\omega)\coth\left(\frac{\hbar\omega}{2 k_B T}\right)\frac{1}{\Omega ^2 - \omega ^2}.
\end{eqnarray}
Choosing $\delta>0$ sufficiently small, and invoking assumption (ii), we can approximate the spectral density around $\Omega$ by an affine function
\begin{equation}
	\omega\mapsto I(\Omega) + C k \Omega^{k-1}(\omega-\Omega),
\end{equation} 
as the local affine approximation of $\omega\mapsto C \omega^k$, for $k\in \rr$.
In this approximation, and approximating in the high temperature limit in which $\coth(\hbar \omega/(2k_B T))$ is approximated by 
$2k_B T/(\hbar\omega)$, the limit $t\rightarrow \infty$ can be performed. In this approximation, we get
\begin{align}
\label{powerdependence}
	\frac{1}{2\Omega}&\int_{\Omega -\delta}^{\Omega + \delta}d\omega   I(\omega)\coth\left(\frac{\hbar\omega}{2 k_B T}\right)\left(\frac{1}{\Omega-\omega}\left(1-\cos\left(\left(\Omega-\omega\right)t\right)	\right)\right)\\ \nonumber  & \overset{t\rightarrow \infty}{\longrightarrow}\frac{I\left(\Omega\right)}{\beta\hbar\Omega ^2}\left(1-k\right)\ln\left(\frac{\Omega+\delta}{\Omega -\delta}\right).
\end{align}
The other contribution of $D_{xp}^{\rm res}$ depends negligibly on the local power law and only on the value of the spectral density at $\Omega$ itself,
\begin{align}
	&\frac{1}{2\Omega}\int_{\Omega-\delta}^{\Omega+\delta}d\omega \frac{ I(\omega)\coth\left(\hbar \omega/(2k_B T ) 
	\right)}{\Omega+\omega}=\frac{I\left(\Omega\right)}{\beta\hbar\Omega ^2}\times \\
	&\bigg( \ln\left(\frac{\left(\Omega-\delta/2\right)\left(\Omega+\delta\right)}{\left(\Omega+\delta/2\right)\left(\Omega-\delta\right)}\right)+k  \underset{\rm negligible}{\underbrace{ \ln\left(\frac{\left(\Omega+\delta/	2\right)^{2}}{\left(\Omega-\delta/2\right)^{2}}\frac{\left(\Omega-\delta\right)}{\left(\Omega+\delta\right)}\right)}}\bigg).\nonumber
\end{align}
The second term is several orders of magnitude smaller than the first for parameters similar to the ones of the present experiment, and is hence negligible.
Up to this point, the assumptions (i)-(ii) as well as the weak coupling and high temperature limits have been invoked.}

\je{So far we have shown that the dependence on the affine function locally approximating a power law
$\omega\mapsto C \omega^k$, is essentially of the form $D_{xp}=a+b(1-k)$, with
\begin{align}
	a&=D_{xp}^{\rm off}(\infty)+\frac{I\left(\Omega\right)}{\beta\hbar\Omega ^2}
	{{\ln\left(\frac{\left(\Omega-\delta/2\right)\left(\Omega+\delta\right)}{\left(\Omega+\delta/2\right)\left(\Omega-	\delta\right)}\right)}},\\
	b&=\frac{I\left(\Omega\right)}{\beta\hbar\Omega ^2}\ln\left(\frac{\Omega+\delta}{\Omega -\delta}\right).
\end{align}
\je{
As it turns out, the only negative contribution to $a$ comes from the term
\begin{align}
\int _{\Omega +\delta}^{\infty}d\omega  I(\omega)\coth\left(\frac{\hbar\omega}{2 k_B T}\right)\frac{1}{\Omega ^2 - \omega ^2} .
\end{align}
For values $k<0$ and not too oscillatory spectral density $I$, one can safely assume that $a>0$. This is a very mild, but strictly speaking an additional assumption about the spectral density being not
too oscillatory outside the relevant window $U_\Omega$.}}
\je{In conclusion this shows that then 
\begin{equation}
	D_{xp}\geq \frac{I\left(\Omega\right)}{\beta\hbar\Omega ^2}\left(1-k\right)\ln\left(\frac{\Omega+\delta}{\Omega -\delta}\right).
\end{equation}
We are finally in the position to have an expression at hand 
from which we can readily read off a lower bound to $\xi$, the measure of non-Markovianity at stationarity for long times $t\rightarrow \infty$, depending only on
measurable quantities. One finds that the $\mu$ that defined the measure of non-Markovianity $\xi$ as in Eq.\ (\ref{xi}) is bounded from below by
\begin{align}
	\mu \geq  \frac{4}{\pi ^2}(1-k)^2 \ln\left(\frac{\Omega+\delta}{\Omega -\delta}\right)^2 ;
\end{align}
it is straightforward to see that a bound to $\mu$ also gives rise to a bound to $\xi$.
This is significantly larger than zero and orders of magnitudes larger than the value for $k=1$, corresponding to the Ohmic case. 
In this sense, we can make a precise and quantitative link between a local deviation from the Ohmic case $k=1$ and the
resulting non-Markovian dynamics. Often, it is read in the literature that the Ohmic case corresponds to Markovian dynamics: Here, this 
connection is made quantitative.}

\je{To further strengthen this point, we make a model for a spectral density of
\begin{equation}
	I(\omega) = \left\{
	\begin{array}{ll}
		\frac{I(\Omega)}{\Omega}\omega ,\, & \omega\in [0,\Omega-\delta),\\
		\frac{I(\Omega)}{\Omega^k}\omega^k,\, &\omega\in [\Omega-\delta,\Omega+\delta),\\
		\frac{I(\Omega)}{\Omega}\omega ,\, & \omega\in  [\Omega+\delta,\Lambda),\\
		0,\, & \omega\in [\Lambda,\infty).
	\end{array}
	\right.
\end{equation}
for $k\in \rr$, $\delta>0$, and cut-off frequency $\Lambda=10^7\Omega$. For this generic spectral density 
\begin{equation}
	D_{xp} = \frac{I\left(\Omega\right)}{\beta\hbar\Omega ^2}\left(1-k\right)\ln\left(\frac{\Omega+\delta}{\Omega -\delta}\right),
\end{equation}
up to a term merely logarithmically divergent in $\Lambda$ 
{which is negligibly small for reasonable cut-off frequencies }.}
\je{
 Then, $k=1$ precisely corresponds to a negligible 
measure of non-Markovianity, that is, to Markovian dynamics. Having said that, the link between spectral densities that can locally
be approximated by power laws or affine functions and non-Markovian dynamics is more generally valid, as pointed out above.}

\subsection*{\jens{Supplementary note 5:\\Impact of technical noise}}

Several potential sources of technical noise in our experimental setup exist that could in principle influence the frequency dependence of the observed noise-floor and might have an impact on our experimental results. The emission from the laser itself, as well as the electro-optical modulator and the photo-detectors could introduce excess, frequency dependent noise, which however can easily be verified experimentally by measuring the light spectrum of the reflected laser field from an unlocked, far off-resonant opto-mechanical cavity. In the frequency band of interest, the noise floor is more than one order of magnitude below the thermal noise floor of the mechanical resonator and shows negligible frequency dependence, which is reflected in our experimental error bars.
{We further rule out noise from our opto-mechanical cavity, by measuring the same cavity without a mechanical element (see Fig.~2). This is done by moving to a mirror pad on the chip that is not released. In addition, electronic noise in our detection system plays a negligible role.}

{Finally, higher order mechanical modes could influence the frequency dependence of our measured noise floor. We have simulated our device and find that these modes do not influence the noise floor in the frequency window we are interested in any significant way. Note that any impact from higher order modes would make the coupling to the environment more supra-Ohmic rather than sub-Ohmic, as observed in our experiment.}

\subsection*{\jens{Supplementary note 6:\\ 
Statistical analysis}}

In this section we describe the statistical analysis in more detail. We process raw data in the form of samples from a time series 
$\{\delta Y^{\rm out}(t)\}$ obtained from the homodyning measurement with a $10$~MHz sampling rate. From this, $m=9,000$ batches \je{of data}
are \je{being} formed, each containing $n=100.000$ data points. 
For every such batch, the data are Fourier transformed, to get data in the frequency domain. \je{In this frequency domain,}
the mean of $100$ data sets is considered. In this way, $90$ independently distributed spectra are obtained. 
For each of these $90$ Fourier transforms, the optimal power $k$ in 
the power law
\begin{equation}
	\omega\mapsto C\omega^k
\end{equation}
is being determined by fixing an interval ${[}\omega_{\rm min}, \omega_{\rm max}{]}$ around the renormalised resonance frequency $\Omega(\infty)$, \je{to minimise the mean square deviation}. 
For the present experiment, 
\begin{eqnarray}
	\omega_{\rm min} &=& 885\text{~kHz},\\
	\omega_{\rm max}&=& 945\text{~kHz} 
\end{eqnarray}
have been chosen. As this minimisation \je{of the mean square deviation} 
constitutes a non-convex optimisation problem, a global simulated annealing algorithm is used. Fig.~3 \je{depicts} 
the outcome of this analysis. Again, while for each realisation of a spectrum several values of $k$ are approximately compatible with the data, one can estimate the optimal $k$ with high significance from the complete data set.

To further corroborate these findings, many variants of the above statistical estimation have been systematically explored. Notably, several instances of bootstrapping, involving a resampling of data, give rise to findings that are indistinguishable from the above ones. The results are largely independent against a different choice of the frequency window or the way batches are being chosen. Also, the findings are not different when not the least squares difference is being considered in the actual intensities, but the logarithms thereof.\\
\end{document}